\documentclass[aps,pre,twocolumn,superscriptaddress,floatfix]{revtex4}
\usepackage{dcolumn}
\usepackage{amsmath}
\usepackage{amssymb}
\usepackage{amsthm}
\usepackage{graphicx}
\usepackage{times}
\usepackage{bm}
\usepackage[T1]{fontenc}
\usepackage{color}
\usepackage{url}
\usepackage[bf]{subfigure}
\usepackage{rotating}
\usepackage{scalefnt}
\usepackage{multirow}

\usepackage{scalefnt}

\theoremstyle{definition}

\newcommand{\W}{\mathbf{W}}
\newcommand{\A}{\mathbf{A}}
\newcommand{\Q}{\mathbf{Q}}
\newcommand{\Pt}{\mathbf{P}}

\newcommand{\vertii}[1]{{\left\vert\kern-0.25ex\left\vert #1 \right\vert\kern-0.25ex\right\vert}}
    
\newcommand{\E}[1]{\left \langle #1 \right \rangle}
\newcommand{\Prob}[1]{\mathbb{P}\left( #1 \right)}

\begin{document}

\title{A recovery rate distribution alters SIS critical properties}
\title{Impact of the distribution of recovery rates on disease spreading in complex networks}

\author{Guilherme Ferraz de Arruda}
\affiliation{ISI Foundation, Via Chisola 5, 10126 Torino, Italy}

\author{Giovanni Petri}
\affiliation{ISI Foundation, Via Chisola 5, 10126 Torino, Italy}

\author{Francisco A. Rodrigues}
\affiliation{Departamento de Matem\'{a}tica Aplicada e Estat\'{i}stica, Instituto de Ci\^{e}ncias Matem\'{a}ticas e de Computa\c{c}\~{a}o, Universidade de S\~{a}o Paulo - Campus de S\~{a}o Carlos, Caixa Postal 668, 13560-970 S\~{a}o Carlos, SP, Brazil.}

\author{Yamir Moreno}
\affiliation{Institute for Biocomputation and Physics of Complex Systems (BIFI) \& Department of Theoretical Physics, University of Zaragoza, 50018 Zaragoza, Spain}
\affiliation{ISI Foundation, Via Chisola 5, 10126 Torino, Italy}

\begin{abstract}
We study a general epidemic model with arbitrary recovery rate distributions. This simple deviation from the standard setup is sufficient to prove that heterogeneity in the dynamical parameters can be as important as the more studied structural heterogeneity. 
Our analytical solution is able to predict the shift in the critical properties induced by heterogeneous recovery rates. 
Additionally, we show that the critical value of infectivity tends to be smaller than the one predicted by quenched mean-field approaches in the homogeneous case and that it can be linked to the variance of the recovery rates. We then illustrate the role of dynamical--structural correlations, which allow for a complete change in the critical behavior. We show that it is possible for a power-law network topology to behave similarly to an homogeneous structure by an appropriate tuning of its recovery rates, and vice versa. Finally, we show how heterogeneity in recovery rates affects the network localization properties of the spreading process.
\end{abstract}

\maketitle

Heterogeneity, whether in the nature of the components or in the pattern of connections, is a key characteristic of complex systems. 
This is particularly evident in the case of the spreading of a disease in a networked population, where the inclusion of structural heterogeneity has long been known to radically change the process' critical behavior ~\cite{Vespignani2001, Mieghem09, Goltsev2012, Arruda2017, Vespignani2015, Arruda2018}. As an illustration, consider two classical contagion models, the Susceptible-Infected-Susceptible (SIS) and the Susceptible-Infected-Recovered (SIR) models, which, when evolving on an homogeneous network, present a non-vanishing critical point~\cite{Vespignani2015, Arruda2018}. However, the introduction of structural heterogeneity --in the form of broad degree distributions of the nodes-- can result in a vanishing critical point~\cite{Vespignani2001, chatterjee2009, Vespignani2015, Arruda2018}. More specifically, in the thermodynamic limit, a divergence of the second moment of the degree distribution~\cite{Vespignani2001, Boguna2013, Vespignani2015, Arruda2018} or a divergence in the maximum degree~\cite{Vespignani2001, Boguna2013, Vespignani2015, Arruda2018} imply a vanishing critical infectivity, which in turn has important practical implications for real-world networks as many display very broad ~\cite{chatterjee2009, Boguna2013,Newman2009} --or even, scale-free-- degree distributions \cite{voitalov2018scale}.
While structural heterogeneity is widely accounted for, heterogeneity in the dynamical parameters has received considerably less attention until recently. Indeed, it was mainly studied for the SIR model. A message passing formalism was proposed in~\cite{Newman2010, Sherborne2018} and an heterogeneous mean-field approach in~\cite{Zhen2017}. In the latter, the authors also performed numerical experiments showing that the population can be more vulnerable in this scenario. More recently, this problem was examined on temporal networks in~\cite{Darbon2018}. Here, we focus on a different type of dynamical heterogeneity by providing the first characterization of the SIS critical point when recovery rates are distributed heterogeneously across the population. This case is empirically relevant because heterogeneous recovery rate distributions have been associated with biological differences between individuals~\cite{Segal2003, Fryer2010}, demographic characteristics~\cite{Dorjee2013} and social differences that result in non-homogeneous access to the health system \cite{Barber2017}. In addition, we consider also the case in which correlations arise between structure and dynamics and we show, analytically and numerically, that such correlations can result in the extreme cases of power-law networks with non-vanishing critical points, and of homogeneous networks with a vanishing critical point. Our results complement previous evidence on the SIR model \cite{Zhen2017} and imply that a proper characterization of the dynamical parameters is of utmost importance not only for a better understanding of spreading processes, but also for many practical applications, such as surveillance, forecasting, and resource management.

\textit{SIS model with heterogeneous recoveries.} 
We start by considering a population composed of $N$ individuals with an arbitrary pattern of connections, which can be represented as a network and is described by its adjacency matrix $\A$, which is usually assumed to be symmetric. Each individual can be in one of two states: (i) infected ($Y_i=1$) or (ii) susceptible ($X_i=1$). Using a Markovian approach, the epidemic process is modeled as a collection of independent Poisson processes. In order to model the spreading of the disease through the network of contacts, for each directed edge, $i \sim j$, emanating from the infected individual $i$, we associate a Poisson process with rate $\lambda_{ij}$, $N^{\lambda_{ij}}(t)$ ($Y_i + X_j \rightarrow Y_i + Y_j$). Additionally, for each infected individual, we associate a Poisson process with rate $\delta_i$, $N^{\delta_i}(t)$, modeling the recovery ($Y_i \rightarrow X_i$). This system is statistically described using the order parameter, $\rho$, and the susceptibility, $\chi$, defined as
\begin{equation}
 \rho = \frac{1}{N}\sum_i^N \E{Y_i}, \hspace{1cm} \chi = \frac{\E{{n_I}^2} - \E{n_I}^2}{\E{n_I}},
\end{equation}
where $n_I$ is the number of infected individuals. Both quantities can be directly estimated using Monte Carlo methods, in particular, the quasi-stationary method (QS)~\cite{Arruda2018} and the Gillespie algorithm~\cite{Arruda2018}, where each of the aforementioned processes are simulated and the state of the nodes is evaluated~\cite{Arruda2018}. 

\begin{figure}[t!]\centering
\includegraphics[width=1\columnwidth]{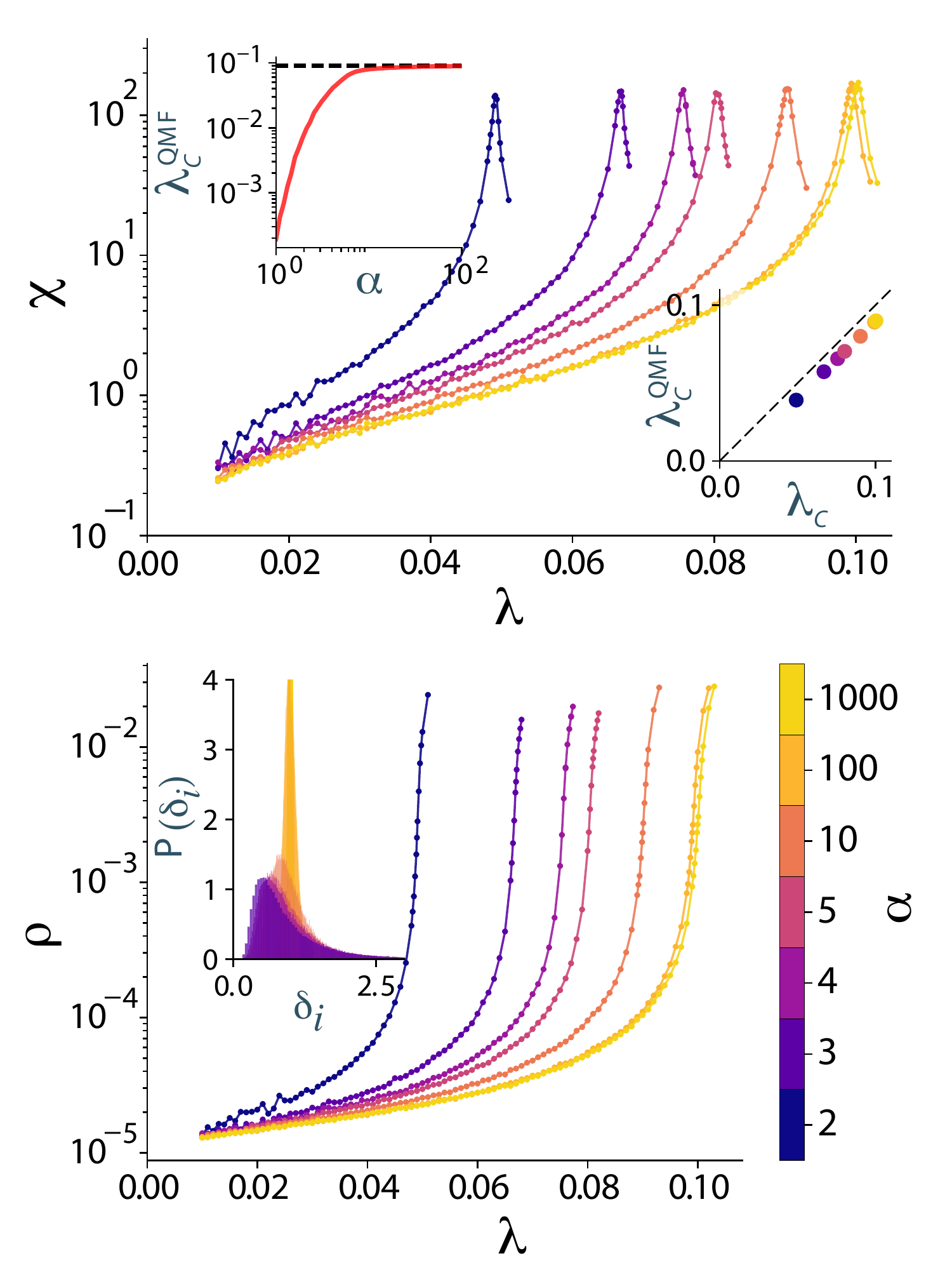}
\caption{Monte Carlo simulations for an Erd\H{o}s-R\'enyi with $N = 10^5$ and $\langle k \rangle \approx 10$ considering that the rate distribution follows a inverse-gamma distribution, whose shaped parameter, $\alpha$, is denoted by the colors. In the top panel, we show the susceptibility curves, the QMF predictions as a function of $\alpha$ in the top inset and the comparison between the QMF estimated and predicted critical points in the bottom inset. In the lower panel, we present the order parameter and the rates distributions in the inset.} 
\label{Fig:InvGamma_distrib}
\end{figure}
\begin{figure}[t!]\centering
\includegraphics[width=1\columnwidth]{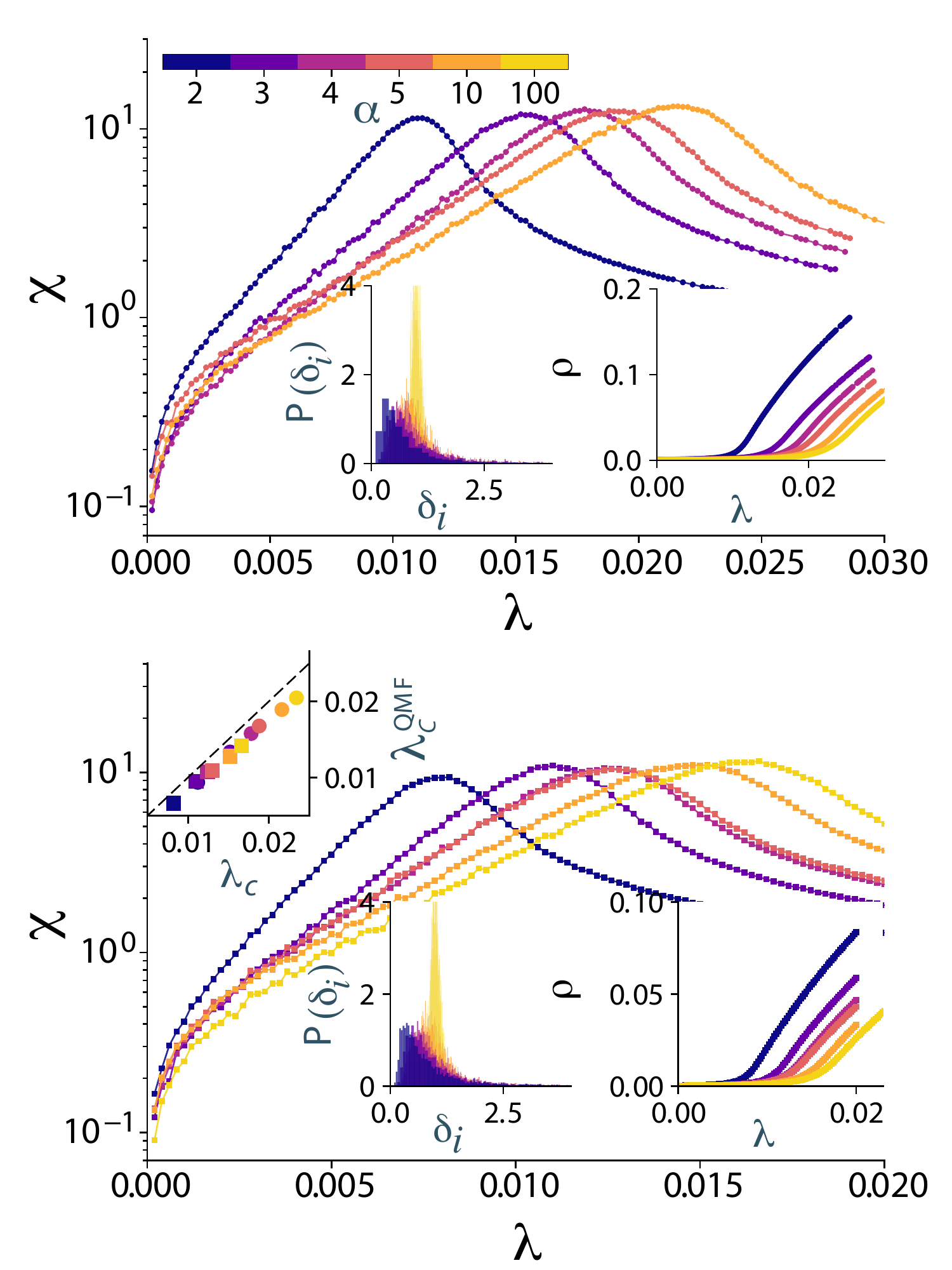}
\caption{In the top panel the \emph{UC Irvine messages} social network \cite{konect:opsahl09, konect} (\textbullet) and in the bottom panel the \emph{openflights} network \cite{konect:2016:openflights, konect} ($\blacksquare$). In both cases, we considered the undirected version of the giant component.  On the main figure of each panel, we present the susceptibility for different values of $\alpha$ and $\lambda$. In the right inset, we present the distribution of recovery rates, which follows an inverse-gamma. In the left inset, we present the order parameter. In the top inset of the bottom panel we present the comparison between the QMF estimated and predicted critical points.}
\label{Fig:Real}
\end{figure}

In the quenched mean field approach (QMF) one implicitly assumes that $\E{X_{i} Y_{j}} \approx \E{X_{i}} \E{Y_{j}}$. 
Physically, this corresponds to neglecting dynamical correlations. Thus, denoting $y_i = \E{Y_{i}}$, we have
\begin{equation} \label{eq:sis_qmf}
 \dfrac{d y_{i}}{dt} = - \delta_i y_{i} + (1 - y_{i}) \sum_{j} \lambda_{ij} \A_{ij} y_{j}.
\end{equation}
The standard SIS model considers that there is no variance in the dynamical parameters, i.e., $\lambda_{ij} = \lambda$ and $\delta_i = \delta$. As a  consequence, it is possible to re-scale time and reduce the parameter space by defining $\tau = \frac{\lambda}{\delta}$. Eq. \ref{eq:sis_qmf} is thus an upper bound \cite{Mieghem09} of $y_{i}$ and, consequently, a lower bound for the critical point, which is calculated as $ \tau^{\mathrm{QMF}}_c = \left( \Lambda_{max} (\A) \right)^{-1}$.
Here $\Lambda_{max} (\A)$ is the leading eigenvalue of the adjacency matrix. 
Note that, for power-law networks (PL), in the thermodynamic limit, the critical point goes to zero if the maximum degree is a growing function of the network size. On the other hand, in the case of a contact process (CP), the spreading rate is defined as $\lambda_{ij} = \frac{\lambda}{k_i}$, and is thus described by the probability transition matrix $\Pt_{ij} = \frac{\A_{ij}}{k_i}$. In this case, the critical point is finite and $ \tau^{\mathrm{QMF,CP}}_c = 1$, regardless of the underlying structure. 

Next, we focus on the case of heterogeneous recovery rates. To this end, consider the most general set-up with heterogeneous parameters allowing an arbitrary distribution of $\delta_i$ and $\lambda_{ij}$. Denoting $\mathbf{y} = \left[ y_i \right]$, near the critical point, we can perform a linear stability analysis of Eq.~\ref{eq:sis_qmf}, hence
\begin{equation}
 \dfrac{d \mathbf{y}}{dt} = \left( \lambda \A \circ \W - \Delta \right) \mathbf{y},
\end{equation}
where $\Delta$ is a diagonal matrix, whose diagonal elements are $\Delta_{ii} = \delta_{i}$, $\lambda_{ij} = \lambda \W_{ij}$ is the rate matrix, and $\A \circ \W$ is the Hadamard product between the rate and the adjacency matrices. 
At the steady state, i.e., when $\dfrac{d \mathbf{y}}{dt} \rightarrow 0$, we have
\begin{equation}
 \mathbf{y} = \lambda \Delta^{-1} \left( \A \circ \W \right) \mathbf{y},
\end{equation}
that will be positive iff the spreading rate is larger than the leading eigenvalue of $\Q = \Delta^{-1} \left( \A \circ \W \right)$, which in turn yields a critical point given as
\begin{equation} \label{eq:threshold_Delta}
 \lambda_c = \left( \Lambda_{max} ( \Q) \right)^{-1}.
\end{equation}
Observe that the elements of $\Q$ are the expected number of contacts before recovery. Obviously, the critical point simplifies to $\tau^{\mathrm{QMF}}_c = \left( \Lambda_{max} (\A) \right)^{-1}$ in the homogeneous case, i.e., when $\delta_i = \delta$ and $\lambda_{ij} = \lambda$. The same applies to the CP. 
Note that, similarly to the homogeneous case, this prediction is an upper bound for the heterogeneous recovery rate scenario, because it relies on the independence of the random variables. In other words, if $i \sim j$, then $\Prob{Y_i = 1 | Y_j = 1} \geq \Prob{Y_i = 1} = y_i$, then the nodal probability is always overestimated (see ~\cite{Mieghem09} for a similar argument). 
From here onward, we fix $\lambda_{ij} = \lambda$ and focus on the effect of the recovery rate distribution on the critical point.  
Note that  Eq.~\ref{eq:threshold_Delta} can be bounded using a matrix norm. Thus, using the 2-norm, we obtain our first result,
\begin{equation} \label{eq:norm_2}
 \frac{\|\A\|_2}{\Lambda_{\max}(\Delta)} \le \|\Delta^{-1} \A\|_2 \le \frac{\|\A\|_2}{\Lambda_{\min}(\Delta)},
\end{equation}
where $\|\A\|_2 = \sqrt{\Lambda_{\max}\left( \A^T \A \right)}$ and, more specifically, $\|\A\|_2 = \Lambda_{\max}\left( \A \right)$ for undirected networks. Eq.~\ref{eq:norm_2} therefore provides bounds on the leading eigenvalue of $\Q$ given by the structure and the variance of $\delta_i$.

\textit{Synthetic networks.} 
To further characterize the critical behavior of our model, we first consider an  Erd\"os -- R\'enyi network (ER) with $N = 10^5$ and $\langle k \rangle \approx 10$ (therefore $\tau_c^{\mathrm{QMF}} \approx 0.1$), which has a homogeneous structure and allows us to analyze the structural and dynamical effects independently. In~\cite{Krylova2013, Clancy2014} the authors showed evidence in real data that the inter-infection time follows a gamma distribution. Consequently, the rate distribution must follow an inverse-gamma distribution. Therefore, we impose the recovery rates to have an inverse-gamma distribution, $\delta \sim \Gamma^{-1} (\alpha, \beta)$,where $\alpha$ and $\beta$ are the shape and scale parameters, respectively. Its mean is $\E{\delta_i} = \frac{\beta}{\alpha - 1}$ and its variance is $\text{Var}(\delta_i) = \frac{\beta^2}{(\alpha - 1)^2(\alpha -2)}$, for $\alpha > 2$. In order to allow the comparison between different distributions, we restrict the distributions to unitary mean. 
In Fig.~\ref{Fig:InvGamma_distrib} we present the critical behavior of an ER network for different shapes, $\alpha$. 
As $\alpha$ decreases, the variance of $\delta$ and, consequently, its maximum, also increases. 
Consistently with Eq. \ref{eq:norm_2}, the critical point also moves toward zero. 
The insets in the top panel emphasize the behavior of the predicted critical point as a function of $\alpha$ and its comparison with the estimations from the Monte Carlo simulations. As expected, for sufficiently large values of $\alpha$ the dynamics behaves similarly to the standard SIS model with uniform $\delta$, where the predicted threshold coincides (see top inset of Fig.~\ref{Fig:InvGamma_distrib}). The agreement between analytical and simulated critical points is very good, as can be seen in the top inset of Fig.~\ref{Fig:InvGamma_distrib}. We remark that a similar experiment was carried out considering a gamma distribution, whose results are presented in Appendix~\ref{sec:A}.

\textit{Real-world networks.} To obtain further insights into real-world epidemics we also consider an inverse-gamma distribution for the recovery rates using a real network. Figure ~\ref{Fig:Real} presents the results of simulations in two real networks: (i) the UC Irvine messages social network~\cite{konect:opsahl09, konect} and (ii) the open flights network~\cite{konect:2016:openflights, konect}. These two networks represent different scales of a similar spreading process: the social network corresponds to a spatially localized network, while the open flights one captures a wider spatial scale. In the top inset of the bottom panel, we observe that the critical point predictions are remarkable for inverse-gamma recovery rates, even for these real networks.

From figures~\ref{Fig:InvGamma_distrib} and~\ref{Fig:Real} we observe that the critical point decreases as we increase the variance of the recovery rate distribution. The critical point predictions for the standard process using QMF are a lower bound. However, when we consider the heterogeneous case, assuming an average recovery rate in the QMF is not enough to provide an adequate characterization of the process. In fact, it is not a lower bound anymore (see Fig.~\ref{Fig:InvGamma_distrib}). The proper correction for the QMF predictions are given by Eq.~\ref{eq:threshold_Delta}, which is a lower bound for the underlying process.

\begin{figure*}[t!]\centering
\includegraphics[width=1\textwidth]{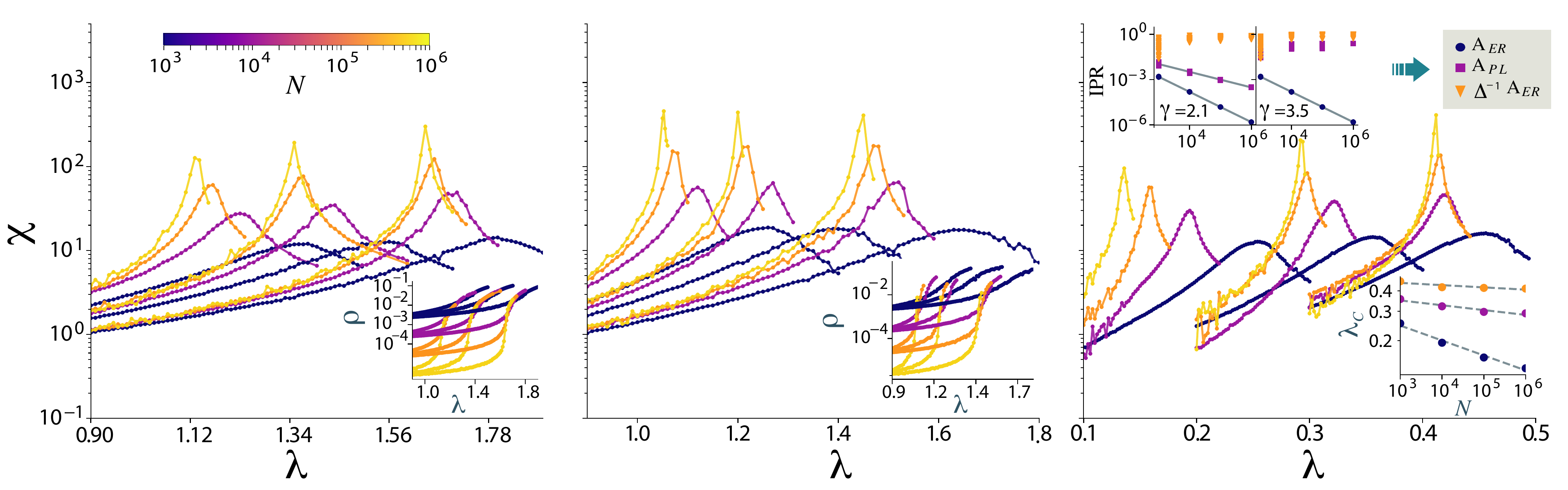}
\caption{Finite size analysis considering structure -- dynamics correlations. In the main panels we present the susceptibility. In (a) the contact process. In (b) SIS with heterogeneous recovery rates, where $\Delta_{ii} = k_i$. Both, (a) and (b) are the same power-law networks with $\gamma = 2.1, 2.7, 3.5$. In the insets we present the order parameter as a function of $\lambda$. In (c) SIS with heterogeneous recovery rates, considering  an Erd\"os -- R\'enyi network and the recovery rates as $\delta_i = \frac{k_i}{k_{PL}}$, where $k_{PL}$ is the same degree distribution as used in (a) and (b). In the inset we present the critical point as a function of the system size in log--log scale. The colors represent the sizes and from left to right, the curves are grouped by the power-law exponents $\gamma = 2.1, 2.7, 3.5$ respectively. In the top insets we show the $IPR$ for the Erd\"os -- R\'enyi, PL networks and the correlated case $\Delta^{-1} \A$, where $\Delta_i = \frac{k_i}{k_{PL}}$ for $\gamma = 2.1$ in the left and $\gamma = 3.5$ in the right. 50 networks were considered in each case.} 
\label{Fig:Cor}
\end{figure*}

\textit{Effects of dynamics-structure correlations.} The bounds in Eq.~\ref{eq:norm_2} implicitly assume that there is no correlations between structure and dynamics. From the Gershgorin circle theorem we know that every eigenvalue of $\Q$ lies at least in one of the disks $D(Q_{ii}, R_i)$, centered in $Q_{ii}$ with a radius given as $R_i = \sum_{i \neq j} |\Q_{ij}|$. Therefore, considering a symmetric matrix, $| \Lambda_k | \leq \Q_{ii} + R_k$, hence $\Lambda_{\max} \leq \| \Q \|_\infty$, where the infinity norm is defined as 
\begin{equation} \label{eq:norm_infty}
 \| \Q \|_\infty = \max_{1 \leq i \leq N} \left( \sum _{j=1}^N \frac{\A_{ij}}{\delta_i} \right) = \max_{1 \leq i \leq N} \left( \frac{k_i}{\delta_i} \right).
\end{equation}
If the structure and the dynamics are uncorrelated, Eq.~\ref{eq:norm_2} is a better bound. However, if they are correlated, Eq.~\ref{eq:norm_infty} might give us further insights. For instance, for the PL case, the leading eigenvalue of $\A$ diverges in the thermodynamic limit leading to a vanishing critical point. Conversely, using Eq.~\ref{eq:norm_infty} and a proper choice of $\delta_i$'s, we can change this behavior. In fact, assuming that $\delta_i (k_i) \propto k_i$ in the thermodynamic, we have
\begin{equation}
 \lim_{N \rightarrow \infty} \| \Q \|_\infty = \lim_{N \rightarrow \infty} \left[ \max_{1 \leq i \leq N} \left( \frac{k_i}{\delta_i} \right) \right] = c,
\end{equation}
where $c < \infty$ is a finite real constant. This radically changes the critical behavior of the dynamics. Note that both the CP and the $\delta_i = k_i$ cases are described, at first order, by the probability transition matrix, $\Pt$, yielding to $\tau_c^{CP} = \lambda_c = 1$. 

In Fig.~\ref{Fig:Cor} we analyze the structure-dynamics correlation effects. In Fig.~\ref{Fig:Cor} (a) and (b) we perform a finite size analysis, comparing both processes on top of PL networks. In (a) we present the CP, where we can already observe that the critical point predictions are not as accurate as for the previous case, in alignment with the predictions reported in~\cite{Ferreira2011, Mata2014}. Thus, the mismatch between prediction and estimated critical point, in this case, seem to be related to dynamical correlations, which is neglected in the QMF. In Fig.~\ref{Fig:Cor} (b) we consider the recovery rate distribution as $\delta_i = k_i$, whose results suggest a finite critical point. However, the convergence seems to be slower if compared with the CP case. 

The previous results show that it is possible, for the same structure, to have a vanishing critical point in the standard model and a non-null critical point when recovery rates are distributed. The opposite scenario is also possible. To show this, we consider an ER network with  $\langle k \rangle \approx 10$ with $\delta_i = \frac{k_i}{k_{PL}}$, where $\Prob{k_{PL}} \sim k_{PL}^{-\gamma}$. That is, we now have a homogeneous structure and a heterogeneous recovery rate distribution. In Fig.~\ref{Fig:Cor} (c) we show a finite size analysis for this configuration varying $\gamma = 2.1, 2.7, 3.5$. We observe that for $\gamma = 2.7$ and $\gamma = 3.5$ the underlying structure plays an important role, maintaining a non-vanishing critical point (see inset of Fig.~\ref{Fig:Cor} (c), where both curves have a slope close to zero). However, for $\gamma = 2.1$ our results indicate the existence of a vanishing critical point (see Fig.~\ref{Fig:Cor} (c) inset). It seems reasonable to hypothesize that the scenario observed when $\gamma = 2.1$ is due to the fact that, in the steady-state, the infection probabilities are inversely proportional to the nodal recovery rate and thus, that the evaluation of the recovery time at both ends of every edge enables an infection-reinfection mechanism. What are the necessary and sufficient conditions to observe this phenomenology needs, however, further exploration. 

Furthermore, note that our model plays an important role in the localization of the leading eigenvector. As shown in~\cite{Satorras2016} and recently formalized in~\cite{Liu2018}, the eigenvector is localized in a sub-extensive portion, i.e., $IPR \sim O(N^{-\nu})$, where $0 < \nu < 1$ and $IPR = \sum_i^N v_i^4$ and $v$ is the normalized leading eigenvector. Thus, in the fully delocalized case, $IPR \sim O(N^{-1})$. As it can be seen in the top inset of Fig.~\ref{Fig:Cor} (c), when heterogeneous recovery rates are considered, the localization of the disease might also change. In one limiting case, the leading eigenvector of $\Pt$ is homogeneously distributed, therefore $IPR \sim O(N^{-1})$ and fully delocalized. However, as shown in the same inset, one can consider a structure that is delocalized for the standard case, but that becomes localized when a recovery rate of the form $\delta_i = \frac{k_i}{k_{PL}}$ (as introduced above) is considered. 
%In addition, note that those are QMF predictions (see top inset of Fig.~\ref{Fig:Cor} (c)). 
We remark that the control of the localization of diseases is still an open problem.

In summary, here we have analyzed the impact of heterogeneity in the recovery rates, allowing it to be arbitrarily distributed. We showed that dynamical heterogeneity is as important as the structural one, and that it can induce drastic changes in the SIS critical properties. Furthermore, an important consequence of our results is that the QMF approach provides a lower bound for the standard SIS, and hence gives a conservative prediction of the critical threshold. However, the standard formulation is not a lower bound for the heterogeneous case, i.e., when assuming $\delta = \E{\delta_i}$ in the classical formulation. To solve this inconsistency, we proposed a solution that relates the structural and dynamical features by the spectral properties of the new matrix $\Q$. Thus, the new formulation opens the path for future research regarding resource allocation, as $\delta_i$ can be associated to the availability of resources to recover individuals. Aside from the specific conclusions drawn here, there are others that concern more general aspects of disease spreading processes as well as the characterization of complex systems in general. For instance, our results might also relate to the predictability of complex systems, and in particular, of diseases \cite{scarpino2017predictability}. At the same time, our findings raise intriguing questions about the consequences of potential heterogeneities in spreading rates, $\lambda_{ij}$ (as suggested by Eq.~\ref{eq:threshold_Delta}), and also in other dynamical processes. For example, in the case of Kuramoto oscillators, correlations between the natural frequencies and node degrees change the nature of the transition. Uncorrelated natural frequencies present a second-order phase transition, while correlations might introduce a first-order phase transition on PL networks~\cite{Moreno2011}. This phenomenology contrasts with what is observed for the SIS model, where, while the phase transition is still second-order, the usual vanishing critical point changes to a well-defined transition.

\section*{Acknowledgement}

Research carried out using the computational resources of the Center for Mathematical Sciences Applied to Industry (CeMEAI) funded by FAPESP (grant 2013/07375-0). Y. M. acknowledges partial support from the Government of Arag\'on, Spain through grant E36-17R, and by MINECO and FEDER funds (grant FIS2017-87519-P).

\bibliographystyle{apsrev}
\bibliography{references}

\appendix
\section{Gamma distributed recovery rates} \label{sec:A}

\begin{figure}[t!]\centering
\includegraphics[width=1\columnwidth]{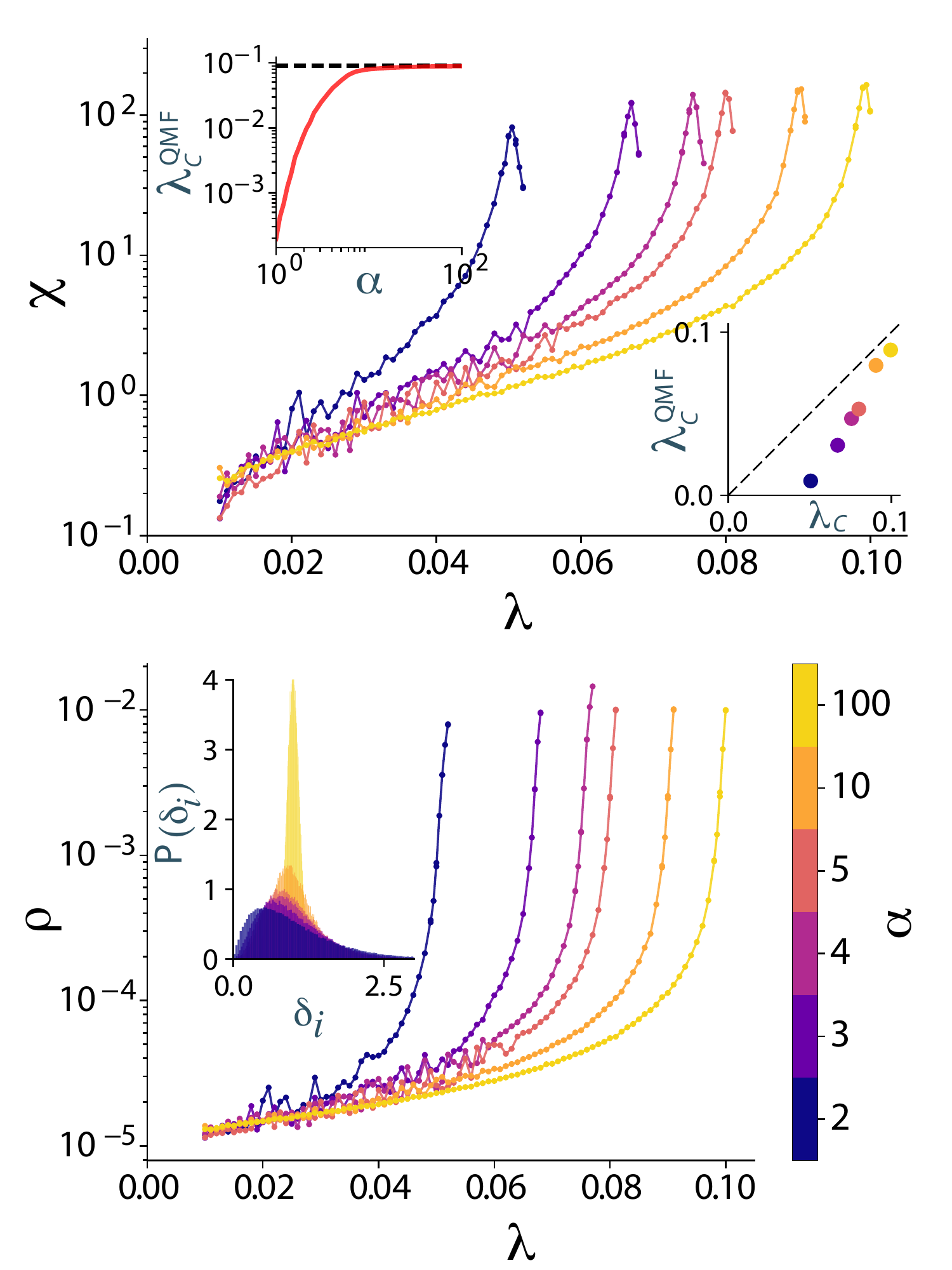}
\caption{Monte Carlo simulations for an Erd\H{o}s-R\'enyi with $N = 10^5$ and $\langle k \rangle \approx 10$ considering that the rate distribution follows a inverse-gamma distribution, whose shaped parameter, $\alpha$, is denoted by the colors. In the top panel, we show the susceptibility curves, the QMF predictions as a function of $\alpha$ in the top inset and the comparison between the QMF estimated and predicted critical points in the bottom inset. In the lower panel, we present the order parameter and the rates distributions in the inset.} 
\label{Fig:Gamma_distrib}
\end{figure}

To further characterize the critical behavior of our model, we first consider an  Erd\"os -- R\'enyi network (ER) with $N = 10^5$ and $\langle k \rangle \approx 10$ (therefore $\tau_c^{\mathrm{QMF}} \approx 0.1$), which has an homogeneous structure and allows us to analyze the structural and dynamical effects independently. 
We impose the recovery rates to have a Gamma distribution, $\delta \sim \Gamma (\alpha, \beta)$, whose p.d.f is expressed as 
\begin{equation}
 f(\delta; \alpha,\beta) =  \frac{\delta^{\alpha-1}e^{-\frac{\delta}{\beta}}}{\beta^\alpha\Gamma(\alpha)}
\end{equation}
where $\alpha$ and $\beta$ are the shape and scale parameters respectively and $\Gamma(\alpha)$ is the gamma function evaluated at $\alpha$. Moreover, its mean is $\E{\delta_i} = \alpha \beta$ and its variance is $\text{Var}(\delta_i) = \alpha \beta^2$. In order to allow the comparison between different distributions, we restrict the distributions to unitary mean by setting $\beta = \alpha^{-1}$ (hence $\text{Var}(\delta_i) = \alpha^{-1}$). 
In Fig.~\ref{Fig:Gamma_distrib} we present the critical behavior of an ER network for different shapes, $\alpha$. 
As $\alpha$ decreases, the variance of $\delta$ and, consequently, its maximum, also increases. 
Consistently with Eq. \ref{eq:norm_2}, the critical point also moves toward zero. 
The insets in the top panel emphasize the behavior of the predicted critical point as a function of $\alpha$ and its comparison with the estimations from the Monte Carlo simulations. As expected, for sufficiently large values of $\alpha$ the dynamics behave similarly to the standard SIS model with uniform $\delta$, where the predicted threshold coincides (see top inset of~\ref{Fig:Gamma_distrib}). Although the agreement between analytical and simulated critical points decreases for very heterogeneous rate distributions, the analytical values for the critical points are always below the simulated ones and thus provide a --safe-- lower bound on the critical threshold.

\end{document}